\documentclass{aa}
\usepackage{graphicx}
\begin{document}

\title{Bendings of radio jets in BL  Lacertae  objects}  
\subtitle{I: EVN and MERLIN observations}  
\author{P.
Cassaro  \inst{1,4} \and C.  Stanghellini  \inst{1} \and D.  Dallacasa  
\inst{3}
\and M.  Bondi \inst{4} \and R.A.  Zappal\`a \inst{4}}

%\offprints{P.Cassaro, \email{cassaro@ira.noto.cnr.it}}

\institute{  Istituto di  Radioastronomia  del CNR, C.P.  141, I-96017  Noto 
SR,
Italy  
\and  
Istituto  di  Radioastronomia  del CNR, Via  Gobetti  101,  I-40129
Bologna, Italy 
\and  
Dipartimento  di Astronomia,  Universit\`a  di Bologna, Via
Ranzani 1, I-40127,  Bologna,  Italy 
\and  
Dipartimento  di Fisica e Astronomia,
Citt\`a Universitaria, via S.  Sofia 78, I-95125 Catania, Italy}

\date{Received / Accepted}

\abstract{ Several  blazars, and  BL Lac objects in particular, show a misalignment  
between the jet  orientation on parsec and kiloparsec  scales.  Some authors  (i.e. 
Conway \&
Murphy,  1993)  have  attempted  to  explain  this  behaviour
invoking helical  jets for misalignment
angles around 90$\degr$, showing how in this case there are interesting
implications for the understanding of the medium into which the jet is expanding. 
By comparing sensitive VLA observations (Cassaro et al., 1999) with images 
available in the literature for the BL Lac objects from the 1-Jy Sample 
(Stickel et al., 1991), it is clear that there is a wide range of misalignments 
between the initial jet direction and the kpc-scale jet, when detected. 
We have carried out VLBI observations of these BL Lac objects, 
in order to 
investigate  the spatial  evolution of the radio jets from few
tens to  hundreds of mas, and to search for helical  jets in this class of 
sources.
We present here the first dataset obtained from EVN+MERLIN observations
at 5 GHz for seven objects.
From these observations we never have
a clear detection of helical jets, we only have a possible signature of their
presence in 2 objects.  In only one of the sources with  a misalignment  angle
around  90$\degr$ the presence of helical jets can be ruled out.
This implies that
it is not  possible to invoke helical jets to explain the morphology of all the 
 sources showing  a misalignment  of about  90$\degr$ between the parsec and the kiloparsec scale jets.
}

\maketitle

\section{Introduction}

While  in  most  radio  sources  it is  possible  to  see
continuity  in the  direction  of the jet, 
a number of blazars shows large misalignment angles between the jet direction 
on parsec and  kiloparsec  scale.  The distribution of the misalignment angles  
($\Delta$PA),  defined as the
angle  between  the parsec  and  kiloparsec  direction  of the radio jet has 
been
studied by several authors  (i.e. Pearson  \&
Readhead,  1988,  Conway \& Murphy,  1993, Hong et al.,  1998).  A 
comprehensive
review on this  argument can be found in Appl et al.  (1996).  They  collect 
the data
on 155 sources from six different blazar samples, obtaining
a misalignment angle distribution ranging from 0$\degr$ to 180$\degr$.  Restricting  the  study to the BL Lac  objects  the  distribution  show 
an
expected peak around  0$\degr$, but also a second peak around  90$\degr$,
harder to explain.  Moreover this  distribution  is lacking   in sources  with 
$20\degr <\Delta PA< 40\degr$.  Hong et al.  (1998),  have  examined  
the
misalignment angles for the blazars selected in the $\gamma$-rays 
by the EGRET experiment, and found that
the BL Lac and  quasars' $\Delta PA$ distributions are different, the 
quasars being
more aligned than the BL Lac objects,  with average  $\Delta$PA  of 21$\degr$
and 99$\degr$, respectively.

Different explanations for the presence of the  secondary
peak have been proposed.  Conway \& Murphy (1993) claim that it would be 
impossible to explain the 
secondary peak through a simple
bend or a double  population  of simply  bent  sources,  proposing that the jet 
have a  helical
trajectory  generated  by  Kelvin-Helmholtz  instability,  probably  induced  
by
precession of the rotation  axis of the central  black hole or of the  
accretion
disk.  Such a mechanism  drives the jet in a fixed helical  trajectory, a sort 
of
tunnel in which the relativistic plasma is flowing.  An alternative is the 
ballistic jet case, in
which  different  components are emitted from the jet basis  following  
straight
paths whose direction  depends on the  instantaneous  direction of the jet 
basis
itself.  In the  case of a  non-ballistic  helical  jet an  interesting  role 
is
played  by the resonance frequency $\omega^{*} \simeq 1.3 c_{\rm ext}/R$ (Conway \& Wrobel,  1995), 
which
depends  on the external (to the jet) sound  speed  $c_{\rm  ext}$ and on the 
transversal jet  radius,  $R$.  This
latter,  indeed,  grows  until  $\omega^{*}$  is  greater  than  the  
precession
frequency, and becomes  constant  (saturation)  when the two frequencies  
become
comparable.  In this phase even the  helix amplitude 
stops growing.
In a FR~II radio galaxy
the intergalactic  medium which 
the jet is going through has a sound speed larger than that required in a
FR~I,  because of the  different  external  density;  moreover  Bridle \& 
Perley
(1984) have determined  that the average  opening angle of the jets of the 
FR~II
is smaller  than that of the FR~I  ones.  These  considerations  lead  them to postulate a greater
resonance  frequency  of the gas in a FR~II  with  respect to that of a FR~I 
and
hence the  saturation  of the helical jet would occur sooner in a FR~I than in 
a FR~II
(Conway \& Wrobel, 1995).  In principle,  this can be useful to distinguish  
the
parent population of the blazar  examined, but without  adequate  statistics
it is  impossible  to use it with some  success. 

A different  mechanism,  based on a distortion of the  accretion  disk, has 
been
proposed  by  Appl  et  al.  (1996)  to explain the misalignment.  They  have  noted  that some 
characteristics
of the misalignments, like the lack of
connection  between the different  scales and locally straight parts of the 
jet,
can be  compatible  with  a  system  of  two  independent  jets,  one  of  
which
originating  in the inner part of the accretion  disk, aligned with its 
rotation
axis, and a second one departing  from the outer part of the disk  having
a different rotation axis due to  Lense-Thirring  precession.  Following 
Appl
et al.  (1996) the  central  jet would be the  mas-scale jet while the  
arcsecond
scale jet would be that  originated  from the outer  region  of the  disk.  
This
model could explain the  independence  of the Doppler factor from the 
$\Delta$PA.
If this is the general scenario, the sources with a small $\Delta$PA 
would be due to the approximate coincidence of the inner and outer axes.

 We observed  seven
BL Lac objects from the 1-Jy sample (Stickel et al.  1991), with the EVN
and the  MERLIN  at 5 GHz to obtain   information  on the  presence  of helical  jets on  blazars  and, if
possible, to trace the path of these jets.  Our aim was to image  the jet from  a few  tens to 
few
hundreds of mas and to study the evolution of the jet between the pc and the 
kpc-scale. Other VLBA data at 1.6 GHz have been obtained and are currently undergoing data processing.

\section{Selection of the sample}

\begin{figure} 
\resizebox{\hsize}{!}{\includegraphics{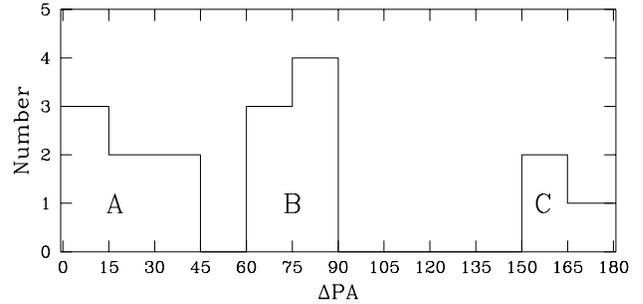}} 
\caption[]{Distribution  of the $\Delta$PA of the 1-Jy sample 
BL Lacs.  It is evident the lack of sources in the intervals 45$\degr$-60$\degr$       and       90$\degr$-150$\degr$.}       
\label{fig1}
\end{figure}

We have chosen to restrict our research to the BL Lac objects of the 1-Jy 
sample
because in a previous study on extended radio luminosity
we have obtained  images for most 
of the sources of this sample with arcsecond  resolution (Cassaro  et al.  
1999).  In order to select the
objects to observe we derived the
distribution  of $\Delta$PA of the sources of the 1-Jy sample, through the
direct inspection of the available images on the arcsecond and the 
milliarcsecond-scales.  
We have been able to determine the $\Delta$PA  for 17 objects only,  either  
because the
lack of information on the pc-scale, or the  difficulty   in  determining a 
clear jet or structure  on the
largest  scale. We did not use  the  values  of
$\Delta$PA  reported in  the literature because they are not obtained with a clearly 
defined
and uniform method. We have compared our VLA images (Cassaro et al. 1999) with 
the 
images on the pc-scale available in literature and those obtained from the 
United States Naval Observatory (USNO) Radio 
Reference Frame Image Database (RRFID, Fey et al., 1996,1997,2000). We have determined the $\Delta PA$ in 
the 
following way: when a clear jet was visible we took the jet
axis as the jet direction in pc-scale images, otherwise we considered the 
straight 
line connecting the core with the  furthest secondary component present in the 
VLBI 
image. We considered as kpc-scale direction of the jet 
the straight line connecting the core and the extended structure (component, 
elongated halo, jet) nearest to the core, which we believe should be the first 
region 
reached by the jet in its path.    In the  distribution  reported in  
Fig.~\ref{fig1} two 
depletions are present:  the first one, already known, between
45$\degr$ and  60$\degr$, and the second one around  120$\degr$, 
less
evident in analogous  distributions  by other  authors (e.g. Appl et al. 1996). 
 Obviously,  this latter
depletion in our distribution  can be due to the poor  statistics.  
We have chosen  objects  belonging  to the three  groups  recognizable in the 
figure, 
that we call A,B, and C according to a $0\degr<\Delta PA<45\degr$, $60\degr<\Delta 
PA<90\degr$  and $150\degr<\Delta PA<180\degr$ respectively,
obtaining the list in Table~\ref{tab1}.

 \begin{table*}  
 \begin{center}  
 \caption{List  of the objects  observed  in this
paper.  Col.  [1] IAU name,  col.  [2] other  name, col. [3] R.A. (J2000) of the pointing center, col. [4] declination (J2000) of the pointing center, col.  [5] optical magnitude (Stickel et al. 1991), col. [6] redshift, col. [7] linear size of 1 mas (pc), col. [8] 5 GHz total  radio flux,    col.    [9]    misalignment    angle     ($\Delta$PA), col. [10] group 
on the misalignment angle distribution (see text).}    
\label{tab1}
\begin{tabular}{ccccccccrc}  
\noalign{\smallskip}  
\hline \hline  
\noalign{\smallskip}
[1] & [2] & [3] & [4] & [5] & [6] & [7] & [8] & [9] & [10]\\ 
IAU name & other name & R.A. (J2000) & $\delta$ (J2000) & m & z & l$_{1mas}$ & S$_{5GHz}$  &  ~~$\Delta$PA & group\\
& & (hh mm ss) & ($\degr$~~$\arcmin$~~$\arcsec$)& & & (pc) & (Jy) & ($\degr$) & \\
\noalign{\smallskip} 
\hline 
\noalign{\smallskip} 
0814+425 &  OJ425 & 08 18 15.999 & 42 22 45.41 & 18.5 & 0.258? & 5.0? & $>$1.69 & 28 & A\\
0954+658 & &  09 58 47.244 & 65 33 54.81 & 16.7 & 0.367 & 6.1 &1.46 & 74 & B\\
1147+245  & OM280 & 11 50 19.212 & 24 17 53.83 & 16.5 & $>$0.2 & $>$4.2 & $>$1.01 & 80 & B\\
1308+326  & AU Cvn & 13 10 28.663 & 32 20 43.78 & 19.0 & 0.997 & 8.5 & 1.53 & 180 & C\\ 
1418+546 & OQ530 & 14 19 46.597 & 54 23 14.78 & 14.5 & 0.152 & 3.4 & 1.09 & 150 & C\\ 
1803+784 & & 18 00 45.683 & 78 28 04.02 & 16.4 & 0.684 & 7.9 & 2.62 & 70 & B\\ 
2131$-$021 &  4C-02.81  & 21 34 10.313 & -01 53 17.28 &  19.0 & 1.285 & 8.6 & 2.12  &  36 & A\\
  \noalign{\smallskip}  \hline  \noalign{\smallskip}
\end{tabular} \end{center} \end{table*}

\section{Observations and data reduction}

The radio observations   were  carried  out on  24 November  1998 at 4.99 
GHz dual polarization,  with an effective bandwidth of 14 MHz,  
using the European VLBI Network (EVN) and the  MERLIN,  with the aim to  obtain 
an
angular resolution  covering  structures with angular size ranging from  tens 
to  hundreds of mas.  The successful EVN 
antennas for this experiment
 were  Effelsberg,  Cambridge,  Jodrell Bank (MK2),  Medicina, Noto, 
Onsala,
and Torun.  The  MERLIN  observation  made  use of the  antennas  of
Defford,  Cambridge,  Knockin,  Darnhall,  Jodrell  Bank (MK2) and Tabley.    
Each source  was observed for approximately one hour, with 
short scans distributed at different hour angle to
improve the {\em uv}  coverage.  The  calibration   was  carried  out 
in a standard way, providing system temperature measurments and gain curves, 
at the
Joint  Institute for VLBI in Europe (JIVE) in Dwingeloo (The  Netherlands),  
and
the final  self-calibration and imaging  was performed at the Istituto
di  Radioastronomia in Noto.  We have used the AIPS  package  for the whole 
data
reduction and analysis.

First, we have obtained the images  for the EVN and MERLIN separately 
and hence, if  useful for our aim, we have  combined the two  datasets to 
obtain an
intermediate  resolution  image.  A small offset ($\leq$10\%) was found between 
the amplitudes of the two datasets, as could be determined by comparing the 
data of the Cambridge-Jodrell baseline, common to both datasets. This offset 
 was corrected accordingly.

\section{Images and comments on individual  sources}

The VLBI images  show jet components in good  agreement  with  previous  
observations.
The MERLIN  images, instead, have detected emission at  a few hundreds of mas from 
the 
core in five out of seven  objects.  

Secondary  components  are very weak if compared to
those on the  arcsecond  structure,  suggesting a high efficiency in carrying 
the energy
at these intermediate scales. In Table~\ref{tab2} we have reported for each source  some characteristics of the components on the EVN, MERLIN and (if any) the combined images: in particular the major and minor axis, the distance and the position angle with respect to the core and the flux density.
The images are  reported  in figures  2 through 19.  The reference for
the arcsecond-scale images of each source is Cassaro et al.  (1999), unless 
otherwise
stated.\\ 
\begin{table*}
\begin{center}
\caption{Data of the components on the images: col. [1] name of the source; col. [2] image data: ``EVN'' for VLBI data, ``MER'' for MERLIN data, ``COM'' for combined data; col. [3] identity letter of the components, as in the figures; col. [4] and [5] major and minor axis of the component (in mas), if pointlike, ``ext'' if extended component; col. [6] position angle; col. [7] distance from the core in mas and col. [8] linear projected distance in pc; col. [9] position angle of the component with respect to the core (counted counterclokwise from the north); col. [10] flux density of the component}
\label{tab2}
\begin{tabular}{cccccrrrrr}
\noalign{\smallskip}
\hline \hline
\noalign{\smallskip}
[1] & [2] & [3] & [4] & [5] & [6] & [7] & [8] & [9] & [10]\\
Name & Data & Comp. & A$_{maj}$ & A$_{min}$ & PA & R$_{core}$ & $l_{proj}$ & PA$_{core}$ &  S$_{5GHz}$\\
 &  &  & (mas) & (mas) & ($\degr$) & (mas) & (pc) & ($\degr$) & (mJy) \\
\noalign{\smallskip}
\hline 
\noalign{\smallskip}
0814+425  & EVN & A & 5.7 & 4.8 & 44 & 0 & 0 & 0 & 961\\
          &     & B & 7.6 & 6.6 & 178 & 3 & 16 & 114 & 67\\
          & MER & A & 48 & 42 & 26 & 0 & 0 & 0 & 899\\
          &     & B & 188.0 & 60.1 & 109 & 97 & 487 & 128 & 12\\
0954+658  & EVN & A & 7.0 & 4.4 & 145 & 0 & 0 & 0 & 250\\
          &     & B & ext & ext & 0 & 12 & 74 & -68 & 21\\
          & MER & A & 48.0 & 37.5 & 93 & 0 & 0 & 0 & 276\\
1147+245  & EVN & A & 11.4 & 4.5 & 49 & 0 & 0 & 0 & 604\\
          &     & B & ext & ext & 0 & 42 & 177 & 102 & 112\\
          & MER & A & 64.3 & 39.2 & 135 & 0 & 0 & 0 & 639\\
          &     & B & 64.0 & 43.8 & 139 & 130 & 549 & -155 & 26\\
          & COM & A & 19.1 & 10.3 & 31 & 0 & 0 & 0 & 605\\
          &     & B & 21.0 & 11.7 & 36 & 11 & 48 & -90 & 58\\
          &     & C & ext & ext & 0 & 137 & 580 & -163 & 7\\
1308+326  & EVN & A & 10.5 & 3.8 & 35 & 0 & 0 & 0 & 2216\\
          &     & B & 13.7 & 8.5 & 54 & 4 & 34 & -90 & 112\\
          & MER & A & 72.7 & 40.0 & 22 & 0 & 0 & 0 & 2106\\
          &     & B & 79.2 & 68.4 & 51 & 90 & 766 & -90 & 14\\
          &     & C & 92.8 & 48.4 & 29 & 230 & 1958 & & \\
1418+546  & EVN & A & 6.5 & 4.1 & 45 & 0 & 0 & 0 & 409\\
          &     & B & ext & ext & 0 & 26 & 89 & 120 & 75\\
          & MER & A & 50.1 & 40.1 & 19 & 0 & 0 & 0 & 353\\
          &     & B & 67.2 & 48.9 & 148 & 20 & 68 & 120 & 29\\
          &     & C & 56.4 & 35.4 & 36 & 130 & 448 & 168 & 2\\
1803+784  & EVN & A & 8.0 & 4.7 & 175 & 0 & 0 & 0 & 1816\\
          &     & B & 8.9 & 8.3 & 77 & 4 & 32 & -92 & 161\\
          &     & C & ext & ext & 0 & 50 & 396 & -132 & 96\\
          & MER & A & 61.4 & 39.2 & 158 & 0 & 0 & 0 & 1879\\
	  &     & B & 71.1 & 59.8 & 122 & 15 & 119 & -81 & 1879\\	
2131$-$021& EVN & A & 13.5 & 4.3 & 40 & 0 & 0 & 0 & 1364\\
          &     & B & ext & ext & 0 & 23 & 198 & 91 & 135\\
          & MER & A & 129.9 & 35.5 & 20 & 0 & 0 & 0 & 1358 \\
          &     & B & 133.8 & 47.3 & 19 & 11 & 95 & 90 & 238\\
          &     & C & 129.4 & 46.0 & 17 & 300 & 2585 & 180 & 8\\
          & COM & A & 26.2 & 10.7 & 18 & 0 & 0 & 0 & 1358\\
          &     & B & 31.0 & 13.8 & 19 & 19 & 164 & & \\
          &     & C & ext & ext & 0 & 138 & 1189 & 163 & 8\\
          &     & D & ext & ext & 0 & 226 & 1947 & 178 & 4\\
          &     & E & ext & ext & 0 & 269 & 2318 & 180 & 2\\
          &     & F & ext & ext & 0 & 320 & 2757 & 175 & 3\\
\noalign{\smallskip}
\hline
\noalign{\smallskip}
\end{tabular}
\end{center}
\end{table*}

\begin{figure} 
\resizebox{\hsize}{!}{\includegraphics{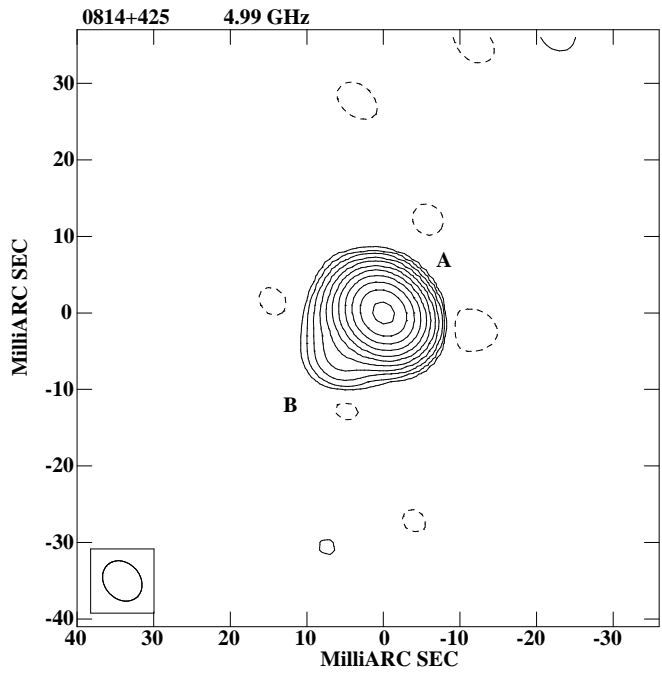}} 
\caption[]{{\bf 0814+425}, EVN image.  The  restoring  beam  is   
5.7$\times$4.6   milliarcsec in PA of 42$\degr$. The noise on the image is  0.25 mJy/beam, the contour levels here and in the following figures are -3, 3$\times 2^i$  (i=0,1,2,...,11) times the rms noise. The peak flux density is 955 mJy/beam.}  \label{0814e}
\end{figure}

\begin{figure} \resizebox{\hsize}{!}{\includegraphics{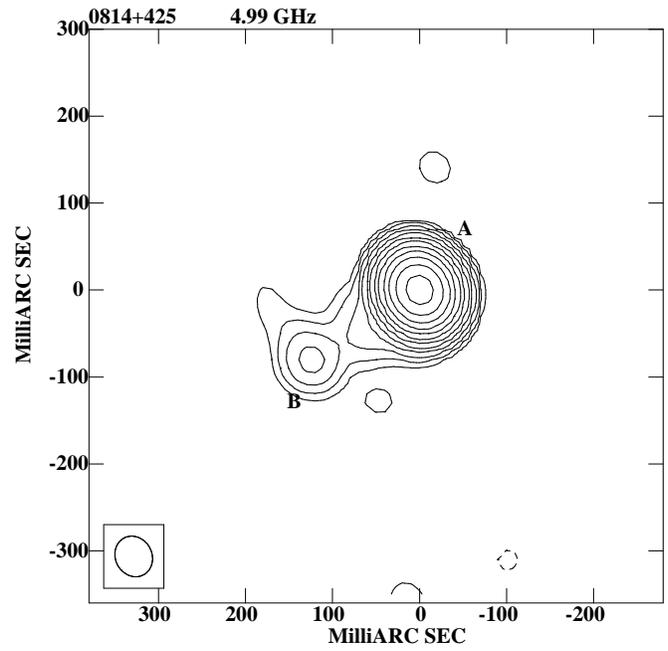}}  
\caption[]{{\bf0814+425},  MERLIN  image.  The  restoring  beam is  
48$\times$42  milliarcsec in PA of 27$\degr$. The noise on the image is  0.1 mJy/beam, the peak flux density is 896 mJy/beam.}
\label{0814m}  
\end{figure}

~\\ 
{\bf 0814+425}:  it is one of the less distorted objects, with a  
$\Delta$PA=28$\degr$.  The VLBI image (Fig.~\ref{0814e}), shows a
hint  of  a  jet  extended  about  10  mas  towards   the SE direction,   confirming   
previous
observations (Polatidis et al., 1995).  The MERLIN image  (Fig.~\ref{0814m})  
shows that the jet 
continues
its path for about  100 mas in a  direction  about 
10$\degr$ bent  towards  the East.  This
structure,  despite  its  initial  bending, is not a sufficient indication
of the presence of a helical  jet.\\

\begin{figure} \resizebox{\hsize}{!}{\includegraphics{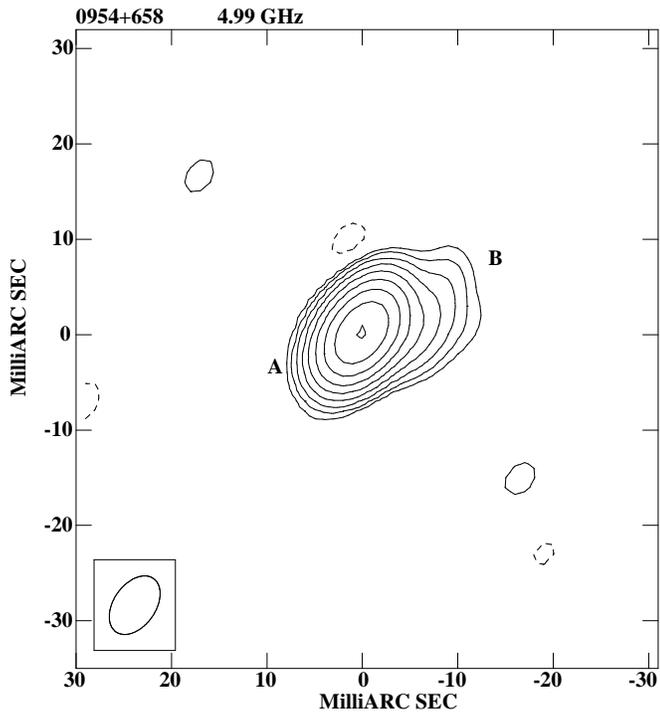}}
\caption[]{{\bf  0954+658},  EVN image.  The  restoring  beam is  
6.8$\times$4.4
milliarcsec in PA of -35$\degr$. The noise on the image is  0.3 mJy/beam, the peak flux density is 241 mJy/beam. }        
\label{0954e}         
\end{figure}

\begin{figure} \resizebox{\hsize}{!}{\includegraphics{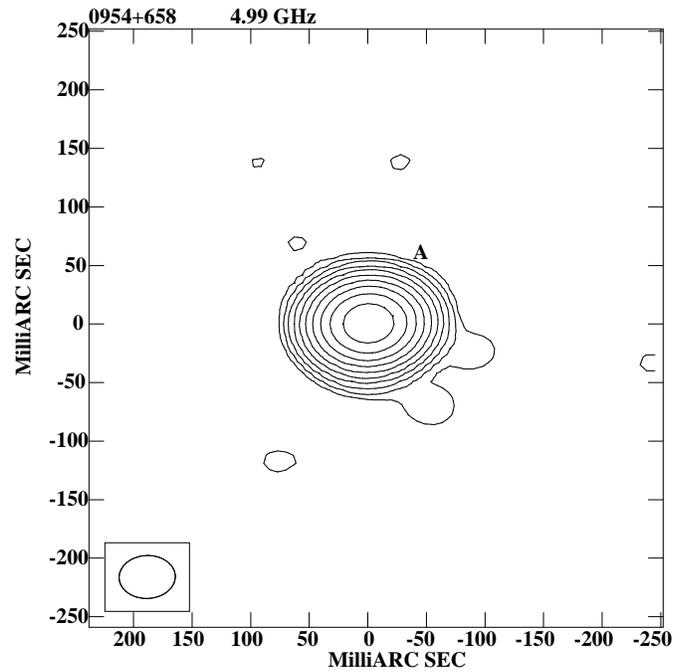}}  
\caption[]{{\bf  0954+658}, MERLIN image.  The
restoring  beam  is   48$\times$36   milliarcsec in PA of -88$\degr$. The noise on the image is  0.1 mJy/beam, the peak flux density is 269 mJy/beam. }  
\label{0954m}   
\end{figure}

~\\
{\bf  0954+658}:  this  object   has  a relatively  small size,  even on the
arcsecond-scale.  With a  $\Delta$PA=74$\degr$ it belongs to the B group
of Fig.~\ref{fig1}.  The VLBI image (Fig.~\ref{0954e})  confirms the 
presence
of a jet in   the W-NW  direction,  slightly  resolved,  but  there  are  no  
evident
components on the   intermediate  scale.  The weak structure  near the core in 
the
MERLIN image  (Fig.~\ref{0954m})  appears not reliable.  

The small
size of the source on the arcsecond-scale and the absence of intermediate  
scale
components lead to a jet slightly bent, with a high $\Delta$PA due to 
projection  effects. The lack of  (visible) jet at the intermediate scale can be 
explained by beaming effects: we see the parsec and kpc-scale jet, while the  
intermediate
scale jet is not detectable because it is highly Doppler boosted away from the
line of sight. \\
   
\begin{figure} \resizebox{\hsize}{!}{\includegraphics{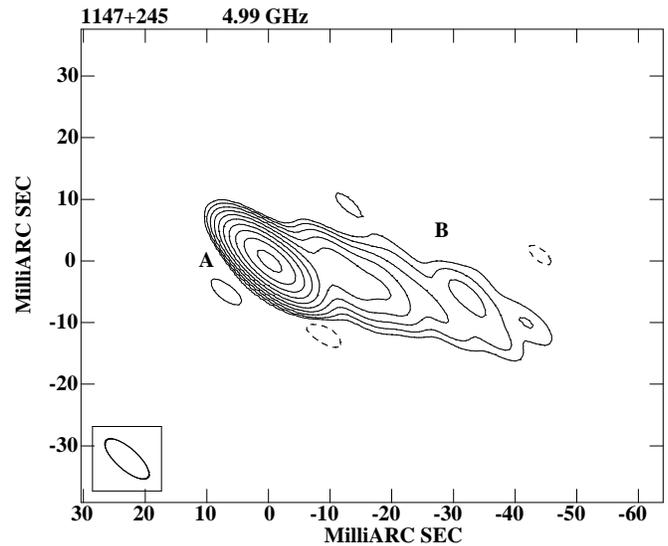}} 
\caption[]{{\bf 1147+245}, EVN image.  The   restoring   beam  is   
9$\times$3.5   milliarcsec in PA of 48$\degr$. The noise on the image is  0.3 mJy/beam, the peak flux density is 576 mJy/beam. }  
\label{1147e}
\end{figure}

\begin{figure} \resizebox{\hsize}{!}{\includegraphics{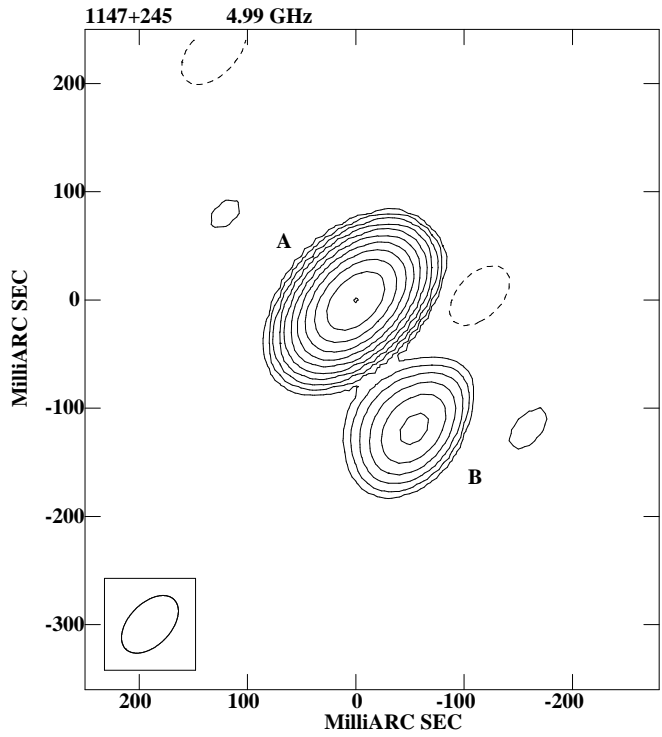}}  
\caption[]{{\bf 1147+245},  MERLIN  image.  The  restoring  beam is  
67$\times$39  milliarcsec in PA of -44$\degr$. The noise on the image is  0.2 mJy/beam, the peak flux density is 630 mJy/beam. }
\label{1147m}   
\end{figure}

\begin{figure}  \resizebox{\hsize}{!}{\includegraphics{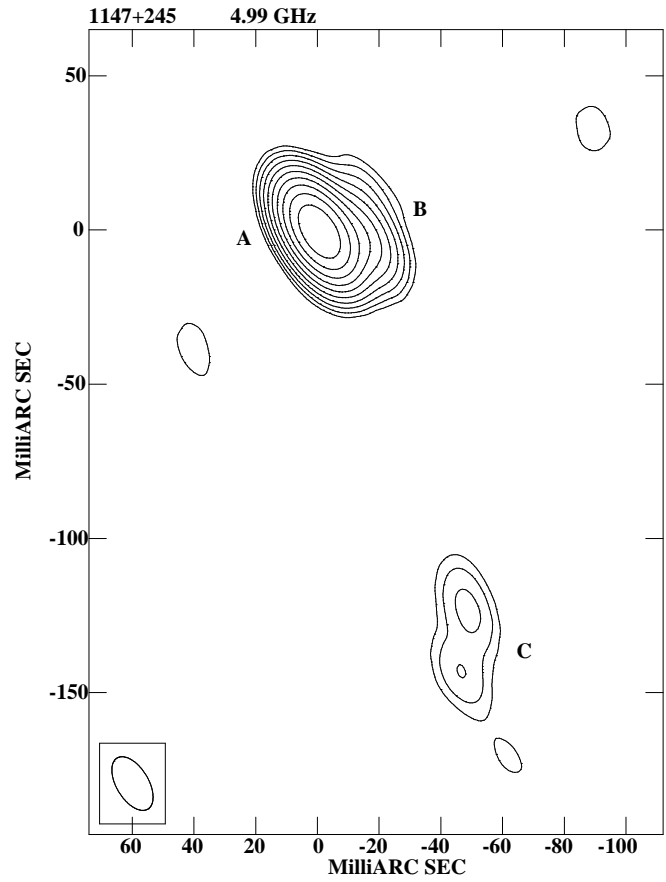}}
\caption[]{{\bf  1147+245},  EVN+MERLIN  combined  image.  The restoring beam 
is 19$\times$10   milliarcsec in PA of 31$\degr$. The noise on the image is  0.2 mJy/beam, the peak flux density is 603 mJy/beam. }   
\label{1147em}   
\end{figure}

~\\  
{\bf   1147+245}:  this   source   has  a
$\Delta$PA=80$\degr$, and the jet at the VLBI  scale  (Fig.~\ref{1147e})  
is
extended about 45 mas towards  the West.
In the MERLIN image  (Fig.~\ref{1147m}) a secondary  component is present at
about 150 mas from the core in the SW direction, at an angle of about  50$\degr$
from the  milliarcsecond-scale jet and  30$\degr$ from the  arcsecond-scale 
jet.  
There is no  trace  of the  bending region between 
the pc-scale jet and the component on the MERLIN scale.  The
 change of direction could be either due to a shock in the interstellar medium
or to the passage  through a region where the magnetic  field  changes  
direction
abruptly  (Koide,  1997), or, finally, to a curvature of the jet 
enhanced  by  projection  effects.  Even if we  cannot  exclude  the  first two
 hypotheses,  we favour the last one,  because of the absence, between the pc 
scale 
jet and the secondary component in the EVN+MERLIN image (Fig.~\ref{1147em}) of 
a bright  region near to the  hypothetical shock in which the jet should dissipate energy and perhaps lose  collimation.  
 In Fig. \ref{1147comb} we present radio images of 1147+245 at different angular scales: from left to right our EVN+MERLIN image, the VLA A-array image at 8.4 GHz obtained by Dallacasa et al. (in preparation) and the VLA A+B array image at 1.36 GHz (Cassaro, 2000).    
The 8.4-GHz VLA image shows that the jet continues to  the South then bends sharply toward  the SE direction. In the 1.36-GHz VLA image we finally see a double structure with wide jets or maybe lobes where the southern emitting region appears  to bend slightly  towards  the SE direction.  
 The structure
is  compatible  with a helical jet and the
abrupt change of direction in the VLBI and VLA 8.4 GHz images would be due to the projection of the helicoid on the
plane of the sky.

\begin{figure} \resizebox{\hsize}{!}{\includegraphics{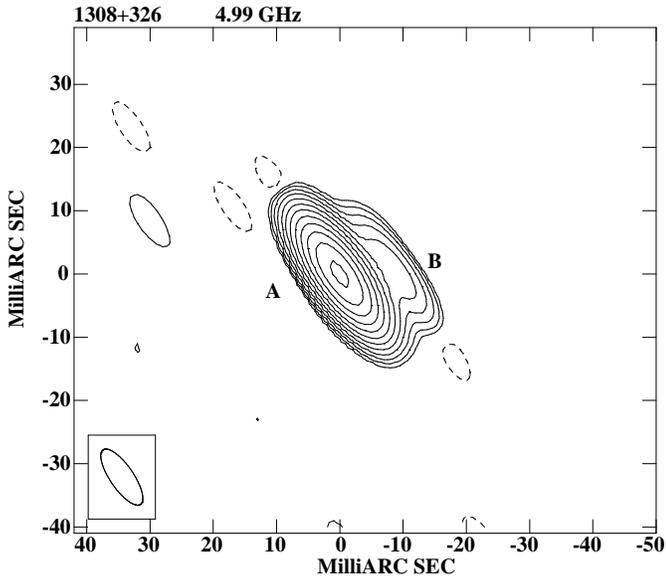}} 
\caption[]{{\bf 1308+326}, EVN image.  The  restoring   beam  is   
10$\times$3.7   milliarcsec in PA of 34$\degr$. The noise on the image is  0.3 mJy/beam, the peak flux density is 2170 mJy/beam. }  
\label{1308e}
\end{figure}

\begin{figure}     \resizebox{\hsize}{!}{\includegraphics{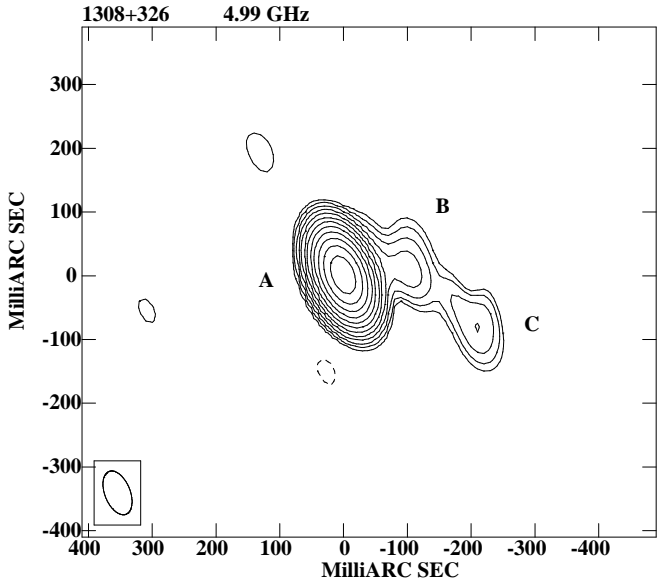}}
\caption[]{{\bf  1308+326},  MERLIN image.  The restoring  beam is  
72$\times$40 milliarcsec in PA of 22$\degr$. The noise on the image is  0.2 mJy/beam, the peak flux density is 2075 mJy/beam. }        
\label{1308m}         
\end{figure}

We note that our images do not cover the scale between  0.15\arcsec and 0.5\arcsec, that would be important to describe the jet behaviour in the first hundreds of parsec, where we expect to better see the jet behaviour. This angular scale should
be adequately covered by the new VLBA data.\\
 ~\\ 
 {\bf  1308+326}: on the  arcsecond
scale this source  shows a diffuse  halo around the core with a luminous  
region
toward  the East and a bright pointlike component at about 10 arcseconds toward  the North (Murphy et
al.  1993).  The VLBI image (Fig.~\ref{1308e}) shows a  jet extended about 10
mas in the W direction,  with a  $\Delta$PA  of  180$\degr$ or 90$\degr$
depending on which arcsecond  structure   we consider as the end of the jet.  
We 
favour the first possibility, given
that the MERLIN image (Fig.~\ref{1308m})  reveals a  well-defined jet which 
departs
to  the West  and after about 100 mas bends  sharply  toward the SW direction, for a total
extension of about 200 mas.  The jet appears to follow a curved path that 
encounters  the  arcsecond-scale East component.  In this  case the  hypothesis 
 of a
curvature due to a shock could be considered, but a helical  structure  can 
also
easily explain the jet trajectory from mas to the arcsecond-scale.  

\begin{figure} \resizebox{\hsize}{!}{\includegraphics{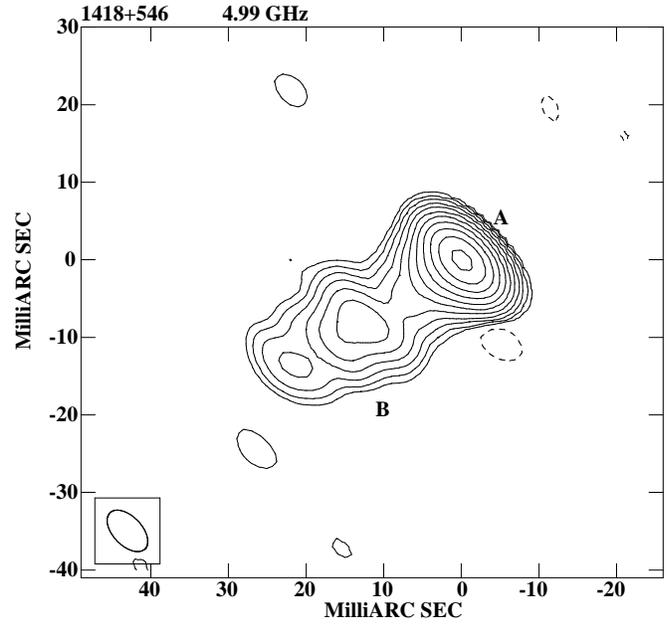}}   
\caption[]{{\bf  1418+546},  EVN  image.  The
restoring  beam  is  6.5$\times$3.7   milliarcsec in PA of 44$\degr$. The noise on the image is  0.1 mJy/beam, the peak flux density is 375 mJy/beam. }  
\label{1418e}  
\end{figure}

\begin{figure}    \resizebox{\hsize}{!}{\includegraphics{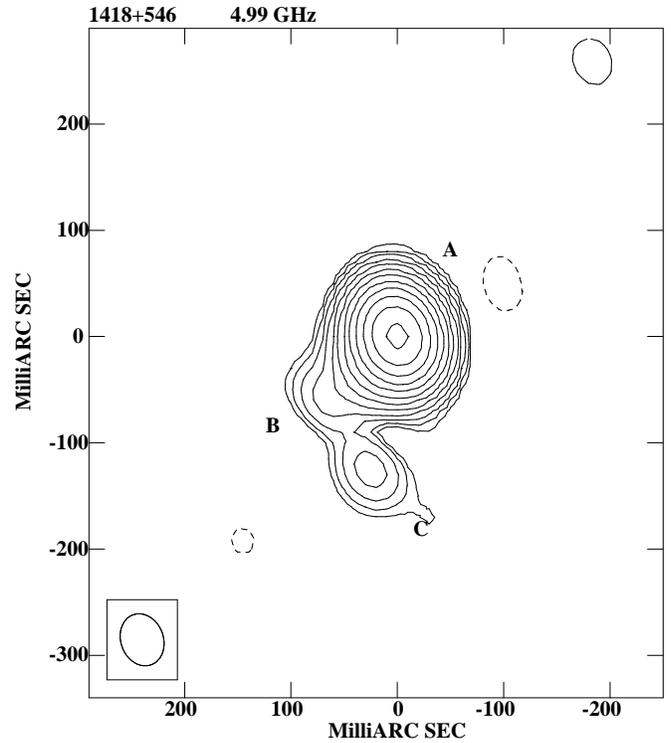}}  
\caption[]{{\bf  1418+546}, MERLIN image.  The restoring beam is  50$\times$41  
milliarcsec in PA of 20$\degr$. The noise on the image is  0.05 mJy/beam, the peak flux density is 364 mJy/beam. }  
\label{1418m}
\end{figure}

\begin{figure}   \resizebox{\hsize}{!}{\includegraphics{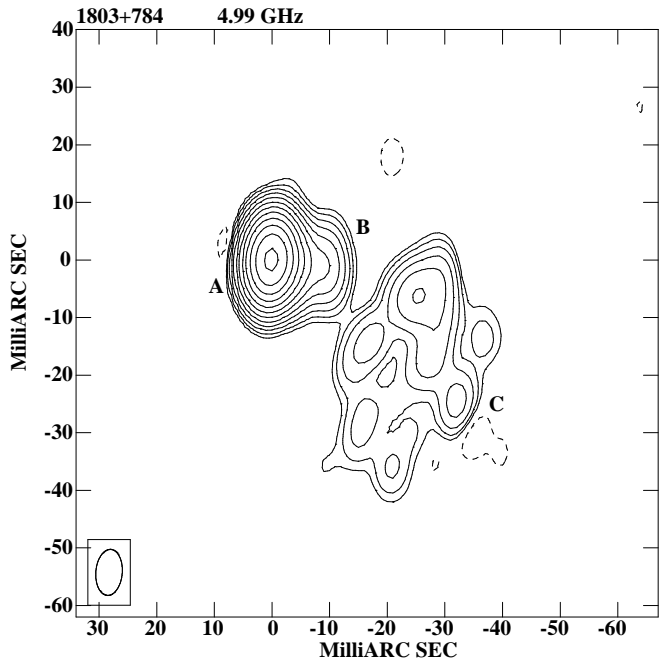}}  
\caption[]{{\bf 1803+784},  EVN  image.  The  restoring  beam  is   
8$\times$4.6   milliarcsec in PA of -5$\degr$. The noise on the image is  0.25 mJy/beam, the peak flux density is 1819 mJy/beam. }
\label{1803e}  
\end{figure}

\begin{figure}    \resizebox{\hsize}{!}{\includegraphics{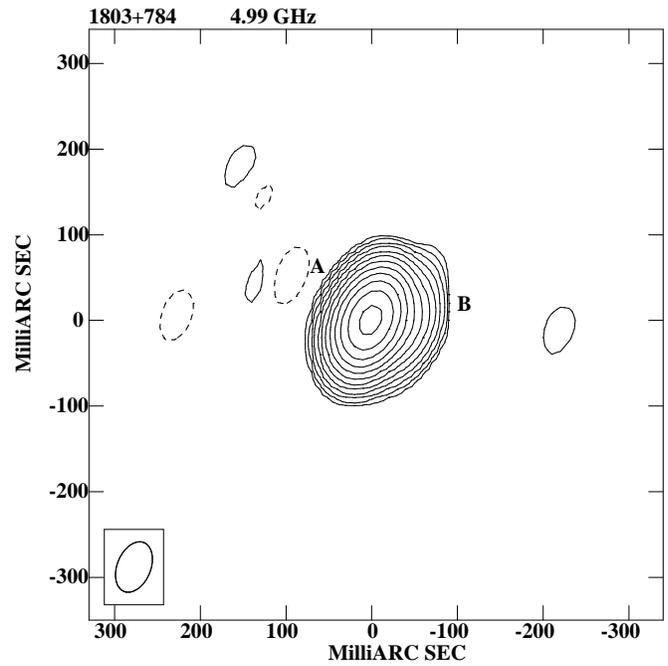}}
\caption[]{{\bf  1803+784},  MERLIN image.  The restoring  beam is  
61$\times$39 milliarcsec in PA of -22$\degr$. The noise on the image is  0.25 mJy/beam, the peak flux density is 1989 mJy/beam. }        
\label{1803m}         
\end{figure}

\begin{figure} \resizebox{\hsize}{!}{\includegraphics{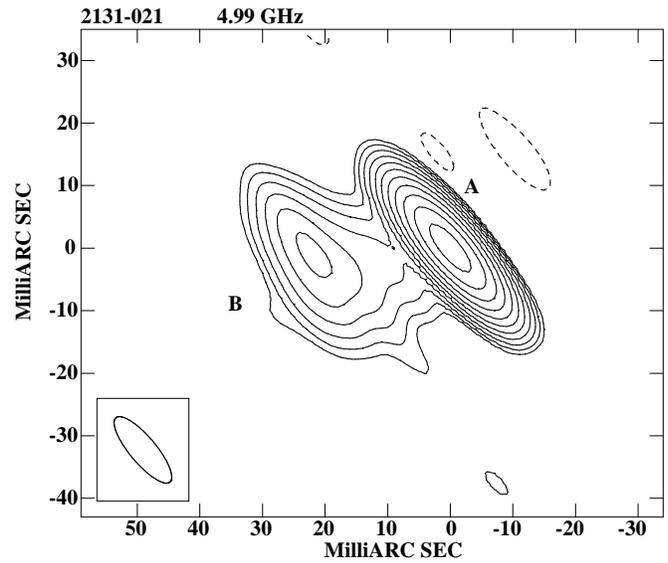}}   
\caption[]{{\bf  2131-021},  EVN  image.  The
restoring  beam  is  13$\times$4.2   milliarcsec in PA of 40$\degr$. The noise on the image is  0.3 mJy/beam, the peak flux density is 1341 mJy/beam. }  
\label{2131e}   
\end{figure}

\begin{figure}   \resizebox{\hsize}{!}{\includegraphics{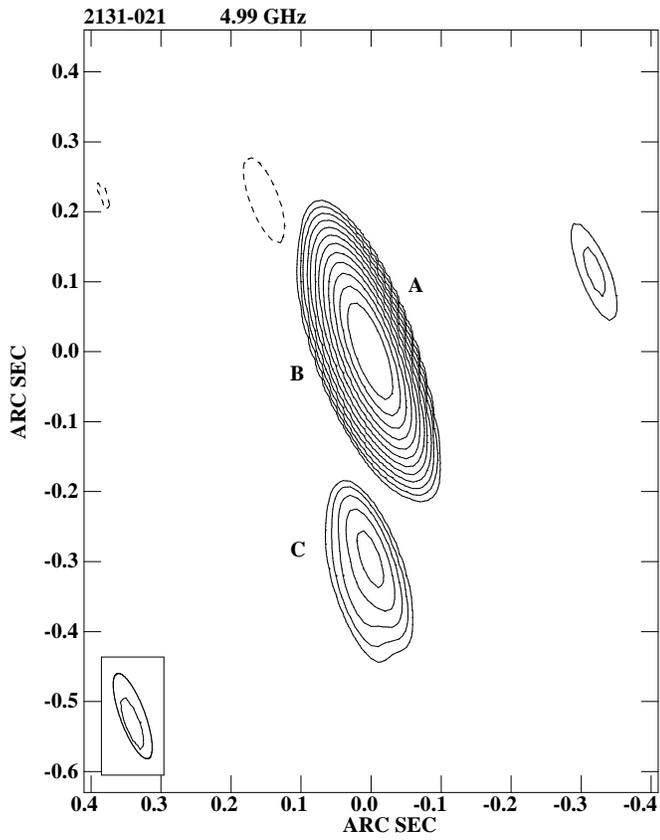}}  
\caption[]{{\bf  2131-021}, MERLIN image.  The restoring beam is 129$\times$35  
milliarcsec in PA of 20$\degr$. The noise on the image is  0.1 mJy/beam, the peak flux density is 1509 mJy/beam. }  
\label{2131m}
\end{figure}

\begin{figure}    \resizebox{\hsize}{!}{\includegraphics{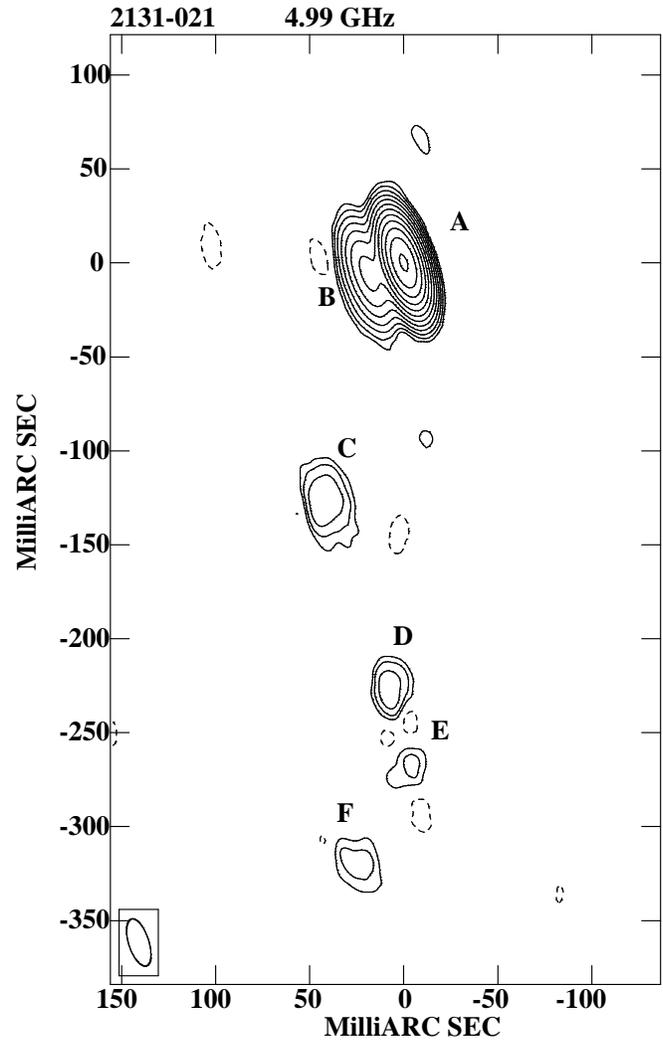}}  
\caption[]{{\bf
2131-021},  EVN+MERLIN  combined  image.  The  restoring  beam  is  
26$\times$11 milliarcsec in PA of 18$\degr$. The noise on the image is  0.2 mJy/beam, the peak flux density is 1349 mJy/beam. }        
\label{2131em}        
\end{figure}

The observed structures  could be also compatible  with a mechanism  similar to 
that
described for 0954+658.  The jet bends smoothly, describing an arc  from
the mas-scale to the  arcsecond-scale,  which for  projection  effects  would 
be
visible  only in the inner part,  becoming  invisible  between  0.2\arcsec and 
10\arcsec 
 in which region the radio emission is boosted away from the line of sight;
then it  again becomes observable at $\sim$10 
arcsecond of
projected  distance from the core at the opposite side of the core, when the 
velocity of
the gas has slowed down and the Doppler factor decreased enough.  To verify 
this scenario  it would
be  important  to image  the  uncovered  range  of  distance  from  0.2\arcsec to about
1\arcsec-2\arcsec, to search  for  possible  components of the jet, which, if the latter 
description is correct, should  not be found.\\  
~\\ 
{\bf  1418+546}:  in this case 
$\Delta$PA=150$\degr$.  The VLBI image (Fig.  \ref{1418e})  shows a 
straight jet directed toward  the SE direction.
On the intermediate scale (Fig.~\ref{1418m}) this source shows a sharp bending
perpendicular to the VLBI jet direction.  For this source, as in 
the previous one, the misalignment and the structure are easily explained with 
a smoothly bending jet, whose curvature is enhanced by the projection.  \\
  ~\\ 
  {\bf  1803+784}:  this
is the largest  object of the 1-Jy sample, with total  projected
linear size of 444 kpc.  In the VLA image  a low-brightness  region is evident at about 9 arcseconds  toward  the SW direction with  respect to the  core; the total radio structure   resembles a  Wide
Angle Tailed (WAT)  source.  Its  $\Delta$PA  is 70$\degr$ and belongs to 
the
second group of Fig.~\ref{fig1}.  The VLBI image (Fig.~\ref{1803e})  shows a 
collimated jet
 about 10 mas long which abruptly loses  collimation   and ends in a
diffuse halo slightly bending toward  the South, in the  direction of the large scale
structure.  The MERLIN image  (Fig.~\ref{1803m}) does not reveal any further 
emission.  The extended structure in the image of Fig.~\ref{1803e} could 
likely be originated  by a  shock  with a  denser  medium,  after  which  the  
jet
loses collimation and changes its direction towards the arcsecond-scale  structure.  
The enhancement of the apparent  bending angle can be due to the projection onto the plane of the sky.  The  presence  of a  helical  jet for  this
source is unlikely.\\ 
 ~\\  
{\bf   2131$-$021}:  we  have   determined  for  this  source  a
$\Delta$PA=36$\degr$,  although in the VLA image  different  structures  
can
be considered the end of the VLBI jet.  
In order to determine  the  misalignment  we have taken into
account  the  arcsecond-scale component closest to the core, which presumably 
will
be reached first.  The  VLBI  image  (Fig.~\ref{2131e})  shows a
component in the East direction extended for about 20 mas.
In the MERLIN image (Fig.~\ref{2131m}) a second component is present toward  the South at about 280 mas from the core.  The  combined
image (Fig.~\ref{2131em}) shows a jet extended for about 30 mas in the SE
direction, and a series of components extending up to 350 mas from the core. The jet appears well  
collimated, and seems to follow a swinging trajectory, indicating a helical motion. The jet seems collimated and the amplitude of the oscillation constant. If the new VLBA data  confirm this behaviour, it would imply that the saturation takes place in the first tens of mas. 

 In Fig. \ref{2131comb} we present radio images of 2131$-$021 at different angular scales: from left to right our EVN+MERLIN image, the VLA A-array image at 8.4 GHz obtained by Dallacasa et al. (in preparation) and the VLA A+B array image at 1.36 GHz (Cassaro, 2000).    
The 8.4 GHz VLA image shows that the jet axis has a $\sim$45$\degr$ change in its direction while the 1.36 GHz image shows a structure reminiscent of a Narrow Angle Tail (NAT) or a Wide Angle Tail (WAT) radio source. The NAT or WAT classification of this source is also consistent with the change in the direction of the jet axis, which likely occurs at the boundary of the host galaxy.   
 
Even in this case the new VLBA data should cover the scale between 0.35\arcsec and about 1\arcsec missing in our images, and would be useful to  properly trace the jet and possibly determine the saturation. Moreover, it could be possible  to verify where the jet axis changes direction and test the NAT/WAT classification.   \\

\section{Discussion and conclusions}

EVN+MERLIN observations at 5 GHz have been carried out to investigate the 
origin of 
the misalignment
between the  arcsecond  and  milliarcsecond-scale  structures  and to  attempt to derive some  information  from the jet properties on 
intermediate scale.

\begin{figure*}  \includegraphics[width=18cm]{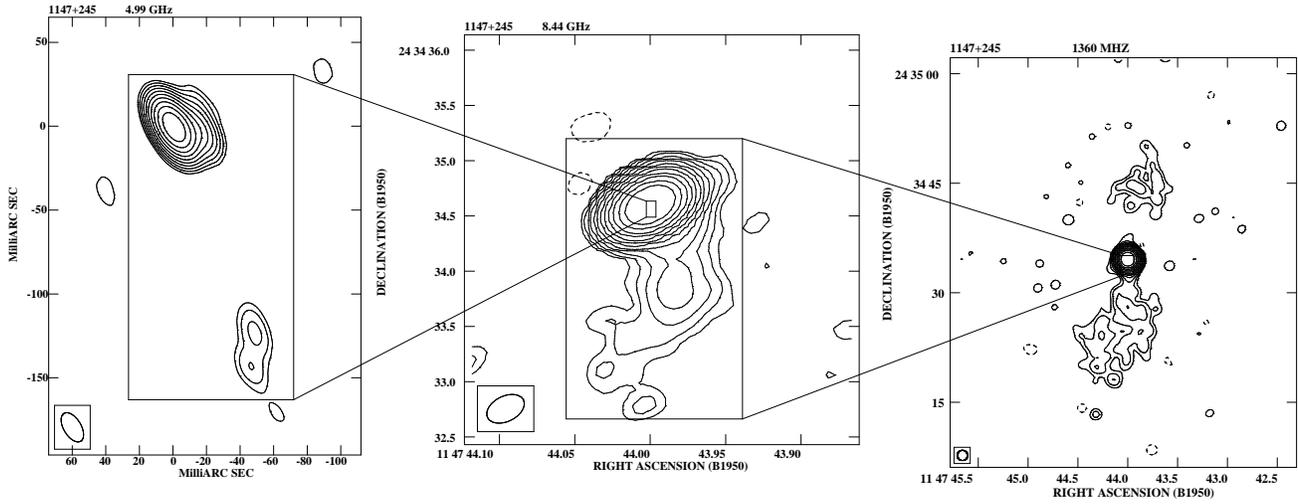}  
\caption[]{Comparison of the images at different scales for the source 1147+245. Left panel: EVN+MERLIN combined image (present paper); center panel: VLA A-array image at 8.4 GHz (Dallacasa et al., in preparation); right panel: VLA A+B array combined image at 1.36 GHz (Cassaro, 2000). }   
\label{1147comb}        
\end{figure*}

\begin{figure*}  \includegraphics[width=18cm]{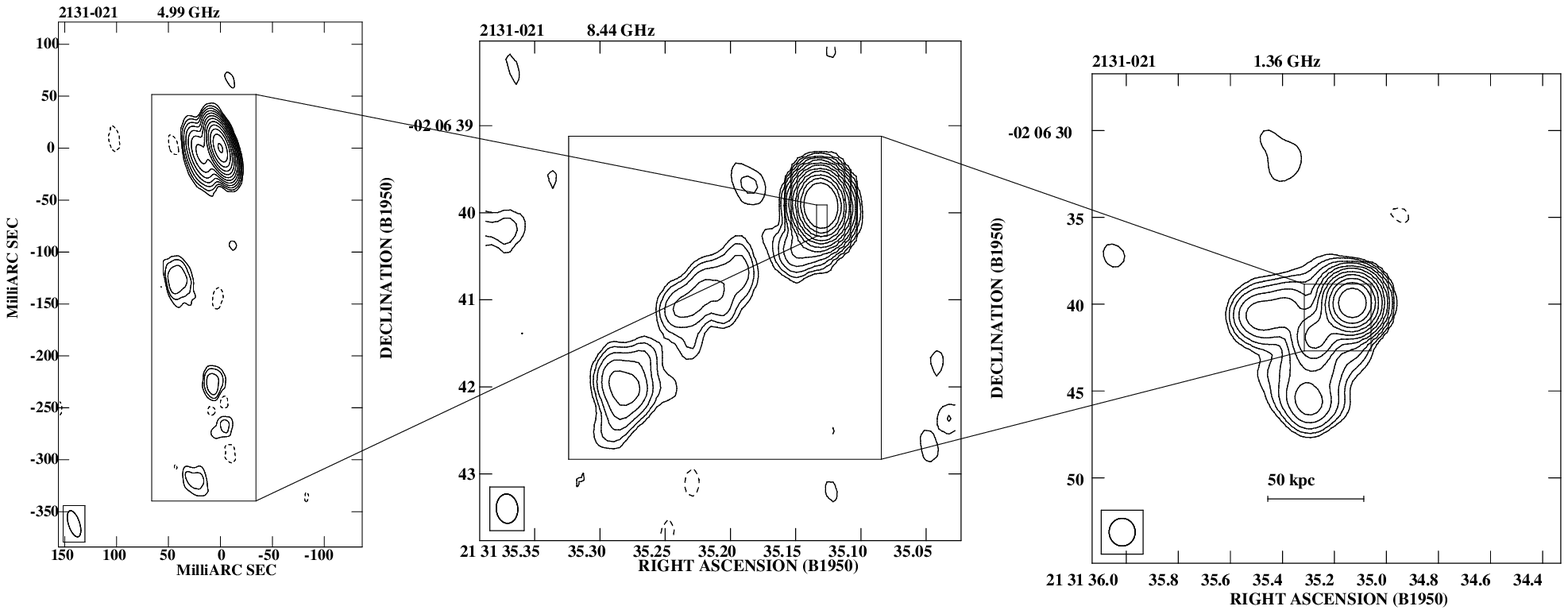}  
\caption[]{The same as in Fig.~\ref{1147comb} for the source 2131$-$021. Left panel: EVN+MERLIN combined image (present paper); center panel: VLA A-array image at 8.4 GHz (Dallacasa et al., in preparation); right panel: VLA A-array VLA image at 1.36 GHz (Cassaro, 2000).}        
\label{2131comb}        
\end{figure*}

A possible  explanation for the misalignment, and in particular for the presence of the secondary peak in the $\Delta$PA  distribution  has been provided by Conway
\& Wrobel (1995) who consider the presence of a helical jet.  Unfortunately, 
the  application  of their
conclusions  has {\it} been only possible for Mrk 501, for which the
images clearly showed the helical structure and the saturation region.  Only 
two sources  observed in this work have morphologies consistent with the presence of a helical jet  (1147+245,
2131$-$021 Figs. \ref{1147comb} and \ref{2131comb}). In no case it has been possible  to clearly  detect 
a helical jet and it
is not possible to determine whether and where the saturation occurs.

Even if there is no sure identification of a helical jet, we can infer some qualitative conclusion from the inspection of our images. The likely helical jets 
 belong to  sources with  different $\Delta$PA: 80$\degr$ for 1147+245, 36$\degr$ for 2131$-$021.  
Moreover, in the radio source 1803+784, the strong loss of collimation favours an
intrinsic bending due to a shock rather than a helical jet.
 In summary, one of the sources
shows a helical jet structure and do not have a 90$\degr$
$\Delta$PA, and at least one other source with a misalignment of about 90$\degr$
does not show a helical structure. This suggests that  Conway \& Murphy's (1993) model, 
which postulates a population of 
helical jets to explain the  90$\degr$ misalignment, may not work
for all objects.

Appl et
al.  (1996) attribute  the peak at 90$\degr$  in the distribution of the 
misalignments
to the presence of two jets  departing
from different  regions of the accretion disk, without considering any intrinsic
distortion of the 
jet.  Several sources  examined in
this work show  characteristics  compatible  with such a  scenario.  Some of
them, however, show components on intermediate  scale  (i.e. 2131$-$021) which 
do not
agree with the presence of only two jets. Moreover some sources have  
intermediate  
scale jets that
begin to bend, as is the case of 1308+326 and 1418+546. 
This is in contrast with the conclusions 
of Appl et al. (1996) who consider straight jets.
We note furthermore that such a mechanism can explain  the 
sources with a misalignment around 0$\degr$, but cannot do the same
with sources belonging to
the secondary peak in the misalignment distributions, or better, it
cannot explain the lack of BL Lac objects with $45$\degr$<\Delta  
PA<60\degr$. 

The misalignment  distribution of Appl et al.  (1996) restricted to the 
quarsars
 lacks the  secondary  peak  around  90$\degr$ which is  present 
instead
for the BL Lac objects. 
This suggests that the mechanism that produces the misalignment is likely to be different
for these two classes of objects.  In particular, for the BL Lacs we can 
conclude
from the images  obtained in the present work, that it is unlikely to invoke a 
single
mechanism to explain the misalignment
between the arcsecond and milliarcsecond
scale structures and the different  structures seen in the intermediate  scale
images.  

While a helicoid is not surely proven to be present in all our misaligned
sources, a less restrictive hypothesis of non-coplanar bending
is required to explain a 90$\degr$ peak in
the distribution.   By non-coplanar
bending we mean a jet for which the directions in starting and ending regions 
of
the path do not lie in the same  plane;  a helicoid itself or a  branch  of a
helicoid resulting from a precession motion of only a fraction of a period, are examples of non-coplanar bending.  
It is also possible to explain
the  misalignment  distribution of the BL Lacs  considering  that  there may
co-exist different kinds of objects. A first category with slightly distorted  jets, and misalignments  distributed from  0$\degr$ to 180$\degr$, with a higher probability of sources with $\Delta$PA around 0$\degr$. 
A second category has  misalignments  around  90$\degr$ caused by non-coplanar jets, helical jets or strong ambient interactions (e.g. 1803+784), explaining the numerical eccess of BL Lac objects with such a misalignment.
\\

\begin{acknowledgements}  
 The European VLBI Network is a joint facility of European and Chinese radio astronomy institutes funded by their national research councils.
MERLIN is operated as a National Facility by the University
of Manchester at Jodrell Bank Observatory on behalf of the UK Particle 
Physics \& Astronomy Research Council. The National
Radio Astronomy Observatory is a facility of the National Science Foundation operated under cooperative agreement by Associated Universities, Inc.

PC acknowledge the Joint Institute for VLBI in Europe (JIVE) for the support and hospitality during the Summer Research Program 1999. PC is also very grateful to Denise Gabuzda and Michael Garrett for the very useful help in data reduction.  

\end{acknowledgements}

%\clearpage           
\clearpage


\begin{thebibliography}{}

\bibitem[Appl  et al.  (1996)]{appl}  Appl S., Sol H.,  Vicente  L., 1996, 
A\&A, 310, 419  
\bibitem[]{}  Begelman M.C., Blandford  R.D., Rees M.J., 1980, Nature,
287,  307  
\bibitem[]{}   Bridle  A.H.,  Perley  R.A.,  1984,  ARA\&A,  22,  319
\bibitem[]{} Cassaro P., Stanghellini C., Bondi M., Dallacasa D., della Ceca R., Zappal\`a R.A., 1999, A\&A, 139, 601 
\bibitem[]{} Cassaro P., 2000, PhD thesis, University of Catania
\bibitem[Conway \& Murphy (1993)]{cm}  Conway J.E., Murphy D.W., 1993, ApJ, 
411,
89 
\bibitem[Conway  \& Wrobel  (1995)]{cw}  Conway J.E., Wrobel J.M., 1995, ApJ,
439, 98  
\bibitem[]{} Fey A. L., Clegg A. W., Fomalont E. B., 1996, ApJS, 105, 299
\bibitem[]{} Fey A. L., Charlot P., 1997, ApJS, 111, 95
\bibitem[]{} Fey A. L., Charlot P., 2000, ApJS, 128, 17 
\bibitem[Hong  et al.  (1998)]{hong}  Hong X.Y.,  Jiang  D.R., Shen
Z.Q.,  1998,   A\&A,  330,  L45   
\bibitem[]{}Koide   S.,  1997,  ApJ,  478,  66
\bibitem[]{}Murphy  D.W.,  Browne  W.A.,  Perley  R.A.,  1993,  MNRAS,  264, 
298
\bibitem[Pearson  \& Readhead  (1988)]{pearson}  Pearson T.J., Readhead  
A.C.S.,
1988, ApJ, 328, 114 
\bibitem[]{}Polatidis A.G., Wilkinson P.N., Xu W., Readhead C.S., Pearson T.J., 
Taylor G.B., Vermeulen R.C., 1995, ApJS, 98,1
%\bibitem[]{}  Raiteri C.M., Villata M., Tosti G., et al., 1999, A\&A, 352, 19
\bibitem[]{}Stickel  M., Fried J.W., Kh\"ur H., Padovani P.,
Urry C.M., 1991, ApJ, 374, 431 
%\bibitem[1999]{villata} Villata M., Raiteri C.M.,
%1999, A\&A, 347, 30 
\bibitem[]{}




\end{thebibliography}
\end{document}